\documentclass[epsfig,11pt]{kluwer}    

\newif\ifpdf
\ifx\pdfoutput\undefined
   \pdffalse
\else
   \pdfoutput=1
   \pdftrue
\fi

 \usepackage{graphicx}                    

\begin{document}                                                                                   
\begin{article}
\begin{opening}         
\title{Turbulence in the Star-forming Interstellar Medium:\\
Steps toward Constraining Theories with Observations}
\author{Mark \surname{Heyer}}  
\runningauthor{Mark Heyer \& Ellen Zweibel}
\runningtitle{Turbulence in the Star-forming ISM}
\institute{Department of Astronomy, University of Massachusetts, Amherst}
\author{Ellen \surname{Zweibel}}
\institute{Department of Astronomy, University of Wisconsin, Madison}
\date{}

\begin{abstract}

Increasingly sophisticated observational tools and techniques are now 
being developed for probing
the nature of interstellar turbulence. At the same time, theoretical
advances in understanding the nature of turbulence and its effects
on the structure of the ISM and on star formation are occurring at a rapid
pace, aided in part by numerical simulations. These increased capabilities on
both fronts open new opportunities for strengthening the links between observation
and theory, and for meaningful
comparisons between the two. 

\end{abstract}

\end{opening}           

\section{Introduction}  

Supernova remnants, expanding HII regions, rotational shear, 
and spiral arm shocks contribute to the 
interstellar gaseous 
maelstrom within galaxies.  Even in the high density regime of the interstellar 
medium, where 
molecules have condensed 
and gravity plays an increasingly larger role
in the dynamics, the flow of gas is chaotic.
In the dense, highly localized cores of giant molecular clouds, self-gravity 
may overwhelm the countering internal pressure,  
enabling the generation of newborn 
stars and stellar clusters.
The 
initial conditions of such protostellar regions are likely 
set
by the overlying turbulent gas.  Therefore,
understanding the critical process of star formation in galaxies 
requires more accurate descriptions of interstellar turbulence, especially
as it relates to the formation of molecular clouds and within molecular gas itself.
Such descriptions demand both insightful theories and relevant
observations
that confront and constrain physical models.  

The study of turbulence within the cold, dense interstellar has greatly 
benefited from the interplay between theory (analytical and numerical 
simulations) and observations.
Analytical efforts target
specific physical processes and typically predict
dimensional relationships that may be measured by the observer
(Kolmogorov 1941; Goldreich and Sridhar 1995; Boldyrev 2002, papers by Chandran
and Boldyrev in these Proceedings).
Yet, purely analytical methods do not follow the 
evolutionary state of a turbulent medium without making overly 
simplistic assumptions nor do 
they readily 
account for complex gas 
distributions driven by advection.  

The sophistication and dynamic range of the 
hydrodynamic and magnetohydrodynamic numerical 
simulations of interstellar turbulence have greatly expanded in 
recent years owing to ever increasing computational capabilities.
The simulations 
do follow the time evolution of an interstellar volume 
and provide 
a 3-dimensional spatial view of the gas distribution and kinematics. 
The primary limitations of current simulations are the low  kinetic and magnetic Reynolds numbers 
relative to expected ISM values (Zweibel, Heitsch, \& Fan 2003), together with the impracticality of 
using the most highly resolved, and hence highly resource intensive, computations
to carry out parameter studies.
Nevertheless, the 
fields of interstellar turbulence and dynamics, and their relationship to star formation, currently are largely driven 
by the results from these computational efforts [see Mac Low \& Klessen (2003) for a
 recent review].

Observational capabilities have also increased several fold 
over the last decade.
Millimeter wave interferometers can now routinely mosaic primary 
fields of view to generate high resolution ($<$5$\tt ''$) images of 
molecular line and continuum emissions from 
star forming regions.  This capability enables investigators to 
 define relationships between compact cores
within the field and provide a census of protostellar 
outflows (Testi \& Sargent 1998; Williams, Plambeck, \& Heyer 2003). 
Sensitive focal plane arrays on single dish, filled aperture 
telescopes enable the 
collection of high spatial dynamic range images (see Figure~1).  
These data
reveal both the varying 
textures of molecular line emissions
and environmental connections to the more widely distributed atomic and 
ionized gas components of the ISM.
Moreover, since turbulence couples motions 
on many spatial scales within 
the inertial range, the analyses of wide field, spectroscopic 
images of the molecular gas play an essential role in constraining 
models of turbulent gas flow.
\begin{figure}[hbt]
\includegraphics[width=0.95\hsize]{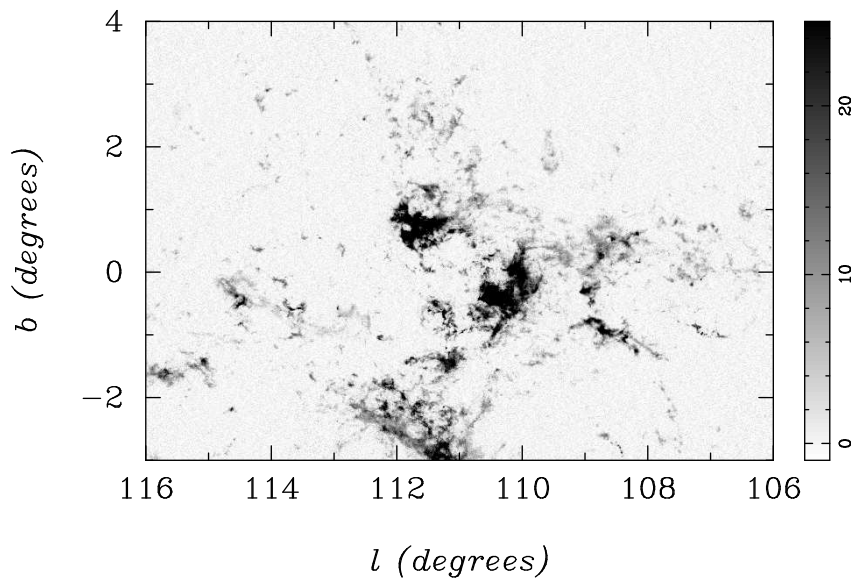}
\caption{An image of velocity integrated CO J=1-0 emission 
from a section of the Perseus 
spiral arm of the Milky Way (Heyer etal 1998).  The image shows giant molecular
clouds, compact globules and flocculent features. }
\end{figure}

In this summary of the 
two sessions on Turbulence in the Interstellar Medium,
we will describe important requirements
in comparing observations with models of turbulence and 
the need for future observations and theoretical advances to make 
progress in this critical field.

\section{A Common Platform}  

A pre-requisite for any useful dialogue between observers and theorists
is a common description of the ISM.  The ISM and the output from 
computational studies are best described by 3-dimensional
scalar (ex. density, temperature) and vector (ex. velocity, magnetism)
{\it fields}.  However, spectroscopic observations do not directly yield such 
3D fields but 
rather, spectra of a given molecular or atomic transition
at different positions in the plane of the sky.
The spectrum is a 
complex integration of scalar and vector field 
quantities along the line of sight and in principle, 
provides a valuable diagnostic to these fields.  In practice,
it is very difficult to reliably extract this information. 
For a turbulent medium, there is a complex re-ordering of the information 
into the spectrum that precludes an 
accurate re-construction of the original fields.
In addition, spectroscopic observations are
subject to selection effects such as line excitation,
opacity, 
and instrumental noise.  These biases act to
mask an uncertain fraction of the volume from the observer
and are not generally 
considered when parameterizing 3D model fields (ex. power spectra of 
the velocity field) or developing predictions.
Since it is not possible to uniquely recover the intrinsic fields 
of density or velocity from the observed spectra, it is necessary to 
compute synthetic spectra from the model fields with the same 
observational biases to achieve a common platform to compare with 
real measurements.
Many investigators have recognized this requirement and have developed 
methods to transform model fields into synthetic spectra taking into 
account radiative transfer and line excitation 
(Falgarone etal 1994; 
Padoan etal 1998; Brunt \& Heyer 2002, Padoan, Goodman, \& Juvela 2003).

Diagnostics of turbulence which are not based on spectroscopy yield valuable
complementary information, but are subject to their own uncertainties. Following
the original suggestion by Chandrasekhar \& M\"unch (1950) that the statistics
of stellar extinction could be used to probe the underlying density distribution
of turbulent interstellar gas, Elmegreen (1997) discussed the distribution of
column densities in a fractal model, and Padoan \& Nordlund
(1999) and Ostriker, Stone, \& Gammie (2001) did the same for numerical models of
supersonic  MHD turbulence. As the latter set of authors have discussed, spatially
and dynamically 
distinct structures 
are frequently projected on top of one another, possibly suggesting an
intrinsic limitation on the use of such statistics for extracting volumetric
quantities.

\section{Calibrated Spatial  Statistics}  

Images of molecular line emission from interstellar clouds exhibit a wide range 
of structural features and textures (see Figure~1).  
The measured spatial variability arises from the various processes that modify 
the gas phase (UV fields from massive stars), compress (shocks from outflows,
HII regions, or turbulence), 
redistribute the gas 
(turbulent eddys), or modulate chemical abundances.
While it is tempting to investigate interesting,
targeted features, such phenomenological efforts 
do not often generate quantitative constraints to theoretical descriptions 
of the ISM. 
Within a 
turbulent medium,
where structure can be generated by advection of material through the velocity 
field, localized, high contrast objects may be short-lived and 
therefore, do not evolve to higher 
density configurations that lead to the formation 
of stars (see Ballesteros-Paredes etal 1999).  

Spatial statistics
provide a condensed, unbiased description of observational and model data
and are an essential tool for the study of interstellar turbulence.  
A frequently used example is the power spectrum of the velocity field,
which is often further parameterized by the slope of the 
power law fit. 
It is essential that any proposed metric derived from the observations 
reflect the true 
statistics of interstellar quantities.  One can not assume that there is 
a direct equivalence given the complexities of the projection described above.
Therefore, it is necessary to calibrate the metric
against fields with known statistics {\it prior} to its application 
on real data.  This calibration 
enables the following results.
\begin{itemize}
\item Demonstrate sensitivity to varying true field statistics.  A derived 
metric that is invariant under varying conditions is simply not 
useful to compare with models.
\item  Empirically define the algebraic relationship between the true 
and measured values as this can be distorted in the projection process.
\item Evaluate the role of noise, resolution, sample size, 
 and selection effects on the results.
\end{itemize}

Several semi-calibrated methods have been developed to estimate the scale 
dependence of the velocity dispersion from spectroscopic data cubes as 
parameterized by the power law slope of the 
second order structure function, or equivalently,
the energy spectrum ($E(k) \sim k^{-\beta}$)  
(Miesch, Scalo, \& Bally 1999, Lazarian \& Pogosyan 2000; Brunt \& Heyer
2002).  However, the accuracy of these methods is not 
yet sufficient ($\sigma_{\beta}\sim0.3$) to even 
fully distinguish between incompressible Kolmogorov turbulence 
($\beta=5/3$) and 
shock dominated Burgers turbulence ($\beta=2$).

In comparison to the use of velocity statistics, diagnostics of turbulence
in molecular clouds based on
magnetic field measurements are in their infancy. Polarimetric maps of star-forming
regions in the far-infrared  probe the orientation of the magnetic field in
regions of high extinction (Hildebrand et. al. 2000). The dispersion of the orientation
angle about the mean contains information about the fieldstrength (Chandrasekhar \& Fermi
1953) as well as the fluctuation spectrum, but corrections for line of sight and angular
averaging can be substantial (Heitsch et al. 2001, Ostriker, Stone, \& Gammie 2001). This
is demonstrated in Figure 2.
\begin{figure}[hbt]
\includegraphics[width=0.95\hsize]{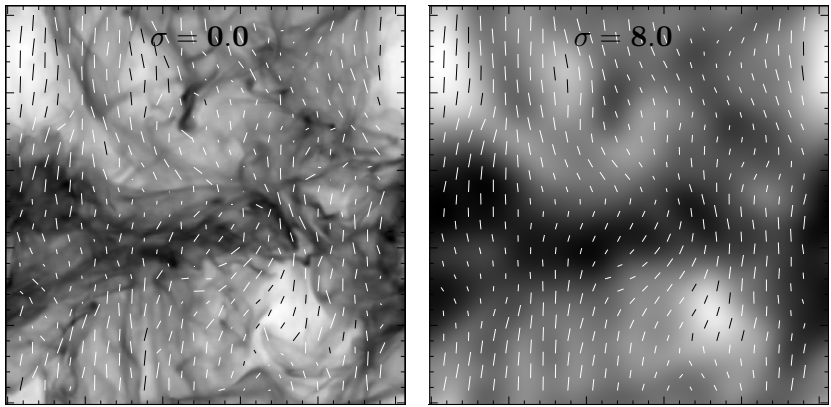}
\caption{Surface density plots and overlaid polarization vectors for a 256$^3$ simulation of
MHD turbulence driven on global scales. The sonic and Alfv\'en Mach numbers are 10 and 3.2,
respectively. The left panel shows unsmoothed data, and the right panel shows the effect of
smoothing with a Gaussian filter with $\sigma=8$ pixels. After Heitsch et al. 2001.}
\end{figure}
Detailed statistical analysis of polarimetric observations of the type emulated in the figure  requires more highly
resolved spatial sampling than is currently available, but would be quite feasible,
for example, with measurements at 100$\mu m$ made by SOFIA.
 
\section{Summary and Future Efforts}  
The discussions that followed the presentations in this session 
focused upon the most essential measurements to conduct in order to constrain 
models and augment
our current understanding of interstellar turbulence.  
\\\\
{\it Mass to  Magnetic Flux Ratio } -- The magnetic field is expected to be frozen to the plasma component of
interstellar gas, except on the smallest scales. Unless the Alfv\'en Mach number of the turbulence is much less
than unity, which is unlikely to be the case, one expects a statistical correlation between fieldstrength and gas density.
Parameterizing the correlation by a power law exponent $\alpha\equiv d\ln{B}/d\ln{\rho}$, one expects $\alpha\sim 2/3$ for
very weak fields, and $\alpha\sim 1/2$ for self gravitating clouds supported by 
moderately strong fields. 
Published observations of the
$B -\rho$ relation suggest that the power law scaling is more an upper envelope than a tight relationship, but the number of
data points is small, and corrections due to line of sight averaging and unresolved 
angular structure could be substantial. But the
absence of a
$B - \rho$ relation, if
confirmed, is evidence for anomalously strong diffusion in the ISM,
 possibly turbulent in nature
(Zweibel 2002, Fatuzzo \& Adams 2002, Kim \& Diamond 2002, Heitsch et al. in these 
Proceedings).

The mass to magnetic flux
ratio $M/\Phi$ is the global counterpart of the density  - fieldstrength relation. 
It measures the
degree to which the evolution of a self gravitating
mass element of the ISM is regulated by its magnetic field. If $M/\Phi$ is above the
critical value $(M/\Phi)_c \sim 0.2G^{-1/2}$, the magnetic field cannot prevent collapse. Measuring the degree of
criticality of dense interstellar clouds is necessary in order to determine the dynamical role of magnetic fields
in star formation. Observations to date suggest that this ratio,
 while not far from critical, generally exceeds the critical value (Crutcher 1999 and these Proceedings,
Bourke et al. 2001, Crutcher, Heiles, \& Troland 2003). 
Possibly this again is evidence for
turbulent diffusion, although these measurements too are subject to small number statistics, line of sight averaging, and resolution
effects.

In order to make further progress, it
is imperative to obtain direct, reliable observations of the magnetic field 
strength. 
In weakly ionized regions this requires measuring Zeeman splitting (difficult), and, in regions with 
substantial electron density, Faraday rotation. Measurements of the polarization of the far infrared continuum and of
radio frequency spectral lines furnish information about the orientation of the field, but not the strength, except indirectly.
Reliable measurements of the gas density are required as well; in too many cases one has only column densities, not volume
densities. Observing more sources, even if only to obtain upper limits, would improve the statistical reliability of the results.
Papers by Crutcher and by Heiles in these Proceedings discuss the situation in molecular and neutral atomic gas, respectively.

It is equally imperative that theoretical models of the $B-\rho$ relation develop observationally testable predictions. For models of
turbulent diffusion, this might involve characterizing the turbulent velocity fields, or predicting the structure function for the
magnetic fieldstrength or the Stokes parameters. Quantifying the effects of numerical diffusion 
is a  separate but equally important issue in simulations of magnetic field evolution
in turbulent media.
\\\\
{\it Degree of Intermittency} - Intermittency describes the filling factor
or sparseness of the spatial and temporal dissipation of turbulent energy.
Signatures to intermittency include non-gaussian probability density 
functions (PDFs) of velocity 
increments (Lis etal 1996; Miesch, Scalo, \& Bally 1999; Pety \& Falgarone
2003) 
and the curvature of the scaling exponent, ${\zeta}(q)$,
with q where 
$$S_q(L) = <|{\delta}v|^q> \propto L^{{\zeta}(q)} $$
is the generalized structure function and ${\delta}v$ refers to the velocity 
difference between two points in a volume separated by size scale, $L$.
It would be useful to 
understand the dissipation process as this is a critical step to the formation 
of stars.   At what scales and through which modes 
is turbulent energy dissipated?  Is this energy replenished? If so,
what are the primary sources of energy injection to 
the interstellar cloud (see Miesch \& Bally 1994)?
The structure function of velocity fluctuations provides a 
statistical road-map of energy 
flow within a volume.  Principal Component Analysis
(Brunt etal 2003) and Velocity Channel Analysis (Lazarian \& Pogasyan 2000)
can recover the low order moments of the structure function.  However,
intermittent phenomena are more readily measured within the 
higher order moments.
The complete evaluation of the generalized structure 
function requires the development of observations and analysis 
tools that are sensitive to 
these higher order ($q > 2$) velocity fluctuations.
\\\\
{\it Density Contrast} -- Collisions of 
super-Alfvenic gas streams produce shocks and 
filamentary density distributions.
Such  
filaments are prominent features 
within the density fields generated by numerical 
simulations of compressible interstellar turbulence.  
The measured angular distribution of molecular line emission is not so 
readily characterized by filaments.  Rather, there are
filaments, compact cores, flocculent clouds, 
and diffuse 
features.  Where are the signatures 
of hydrodynamic shocks?  
Are these lost in the projection of the fields or in the observational 
noise? An important metric to develop is a filamentary index which 
could describe the fraction of mass contained in filaments or sheets
both in 3-dimensional density fields and projected images of gas 
column density.
\\

As stated in the first section of this
article, analytical theories of turbulence in the ISM led to a few testable
predictions, but do not describe the wealth of detail that observations of the ISM
reveal. This made it difficult, in the past, to argue that we understand very basic
problems, such as the nature of energy injection, the primary mechanisms for cloud
formation, the dynamical role of magnetic fields, and the origin of the observed
IMF. Numerical simulations offer a threefold opportunity in this regard. First of all,
numerical simulations can be used to study universal problems in data analysis and
interpretation, such as recovery of 3D distributions from 2D projections, the
role of finite angular resolution, and the role of noise. Studies of this type can
be valuable in the interpretation of the observations, and perhaps can lead to new
methods of data analysis. 
Second, it is possible, through parameter studies,
to determine the sensitivity of the basic output - the density, velocity, and
magnetic fields in a turbulent simulation - to
changes in the basic
physical input. For example, if turbulence is driven at a particular scale, is
that scale imprinted on the output fields? Does it make a difference whether the
driving is through addition of energy, or momentum? What is the effect of underlying
rotational shear? Finally, the first and second avenues of investigation must be
combined to demonstrate the sensitivity, or lack thereof, of observable quantities
to physical processes and input parameters. 
This makes it possible to answer questions such as: What
observations are required to determine whether the turbulence in a molecular cloud
is driven or decaying? Is there a way to determine, short of direct measurement,
whether the magnetic field in a cloud is weak or strong? What is the physical
significance of the observed linewidth - size and column density - size relations?
We are challenged by these and other questions regarding interstellar 
turbulence.  The continued creative tension between observations and 
theory
provides motivation for
future studies
to address these critical issues.  

\acknowledgements

This work is supported by NSF grant AST 01-00793  to the Five College
Radio Astronomy Observatory and NSF grant AST 0098701 to the University of Wisconsin.

\end{article}
\end{document}